\documentclass{webofc}
\usepackage{graphicx}
\usepackage[varg]{txfonts}
\usepackage{sidecap}
\usepackage{braket}
\usepackage{color}
\usepackage{picinpar}
\usepackage{hyperref}

\begin{document}
\title{Towards an electrostatic storage ring for fundamental physics measurements}
%
%

\author{\firstname{C.} \lastname{Brandenstein}\inst{1}\fnsep\thanks{\email{chiara.brandenstein@tum.de}} \and \firstname{S.}
        \lastname{Stelzl}\inst{1}\inst{2}\fnsep\thanks{\email{stefan.stelzl@epfl.ch}}
\and
        \firstname{E.} \lastname{Gutsmiedl}\inst{1}
        \and
        \firstname{W.} \lastname{Schott}\inst{1}
                \and
        \firstname{A.} \lastname{Weiler}\inst{1}
        \and \firstname{P.} \lastname{Fierlinger}\inst{1}
}

\institute{ Technical University of Munich, Physics Department \and  École Polytechnique Fédérale de Lausanne, Theoretical Particle Physics Laboratory}

\rightline{TUM-HEP-1435/22}

\abstract{
  We describe a new table-top electrostatic storage ring concept for $30$~keV polarized ions at frozen spin condition. The device will ultimately be capable of measuring magnetic fields with a resolution of 10$^{-21}$~T with sub-mHz bandwidth. With the possibility to store different kinds of ions or ionic molecules and access to prepare and probe states of the systems using lasers and SQUIDs, it can be used to search for electric dipole moments (EDMs) of electrons and nucleons, as well as axion-like particle 
  dark matter and dark photon dark matter. Its sensitivity potential stems from several hours of storage time, comparably long spin coherence times, and the possibility to trap up to 10$^9$ particles in bunches with possibly different state preparations for differential measurements.
 As a dark matter experiment, it is most sensitive in the mass range of 10$^{-10}$ to 10$^{-19}$~eV, where it can potentially probe couplings orders of magnitude below current and proposed laboratory experiments.
}
\maketitle
\section*{Introduction} 
The axion, a light pseudoscalar particle, has gained increasing attention due to a growing experimental program and new theoretical ideas. The axion could constitute a significant fraction of dark matter. A spin-dependent coupling to ordinary matter then leads to the novel signature of an effective oscillating magnetic field that can be detected using nuclear magnetic resonance techniques, see e.g.,~\cite{Graham_2013,Budker_2014,Garcon:2019inh,JacksonKimball:2017elr}. Related proposals that can test the axion couplings to matter can be found in \cite{Gramolin:2020ict,Arvanitaki:2021wjk,Gao:2022nuq,Lisanti:2021vij,Sikivie:2014lha,Graham:2017ivz,Jiang:2021dby,Chang_2019,Graham_2021,Agrawal:2022wjm,CPEDM:2019nwp,JEDI,Kim:2021pld,Abel_2017,Roussy_2021,Stadnik:2022gaa,Gaul:2020bdq,Berlin:2022mia}, some of which~\cite{Chang_2019,Graham_2021,Agrawal:2022wjm,CPEDM:2019nwp,Kim:2021pld,JEDI} in the context of storage rings. Furthermore, we have yet to find hints for additional sources of CP violation in EDM measurements~\cite{EDMreview}.
	Here, storage ring precision experiments gain increasing attention as they provide alternative access via extreme measurement precision.\\
\noindent
In this paper, we describe a versatile, table-top-scale electrostatic storage ring with 5.7~m circumference for non-relativistic, polarized ions, built to search for electric dipole moments of fundamental systems (EDMs) and to probe magnetic field-like effects resulting from ultra-light dark matter such as axion-like particles (ALPs) and dark photons. Furthermore, a combination of the device with a large volume nuclear spin polarized sample is proposed to increase the sensitivity for the ALP dark matter and dark photon dark matter measurement.
\section{The storage ring}\label{The storage ring}
\begin{figure}[t]
	\centering
	\begin{minipage}[t]{0.45\linewidth}
		\centering
	\includegraphics[width=0.9\linewidth]{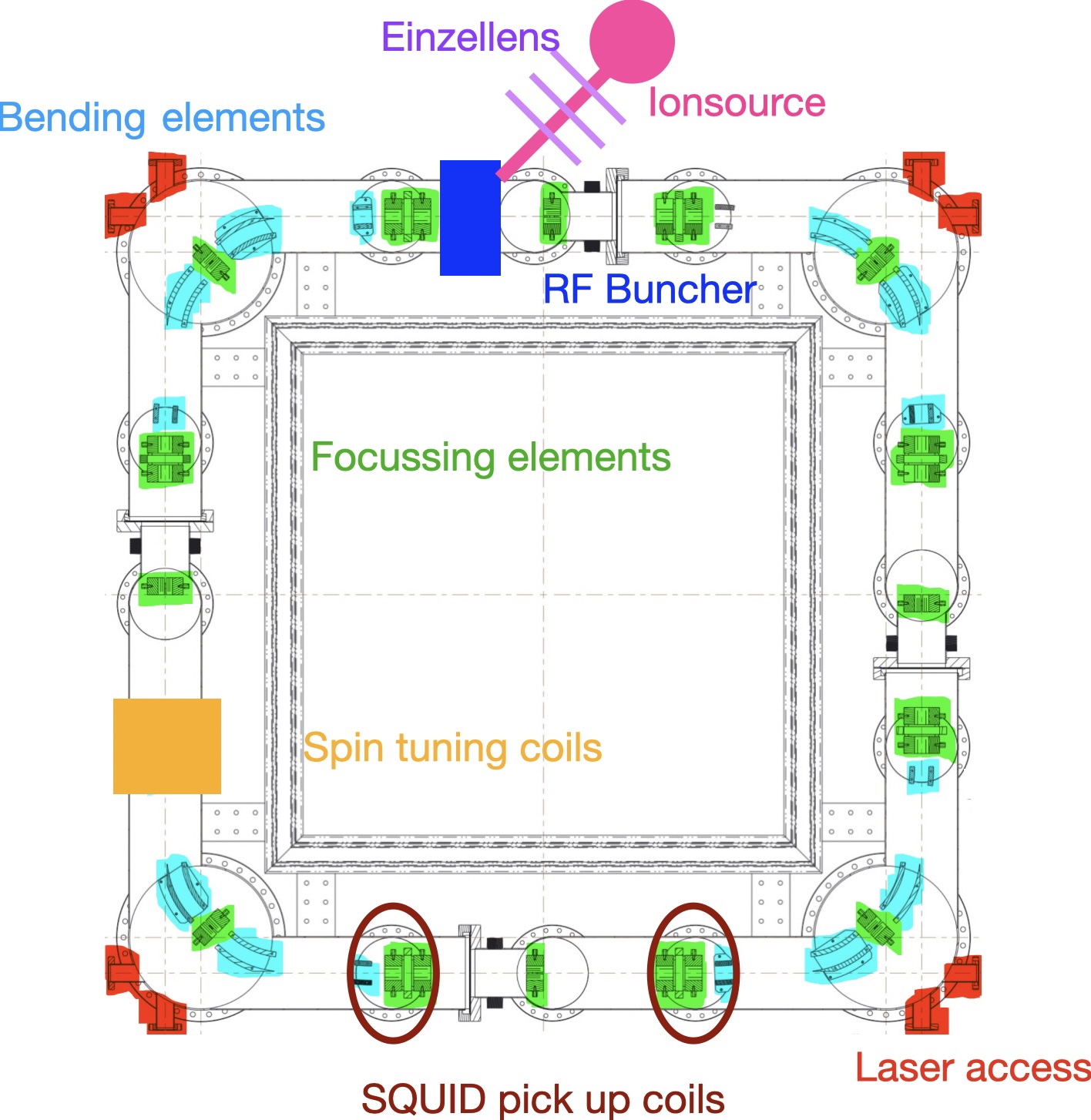}
		\caption{Sketch of the storage ring with bending elements, focussing elements, SQUIDs, spin tuning coils and RF buncher.}
		\label{sketch}
	\end{minipage}
	\hfill
	\begin{minipage}[t]{0.45\linewidth}
		\centering
		\includegraphics[width=0.8\linewidth]{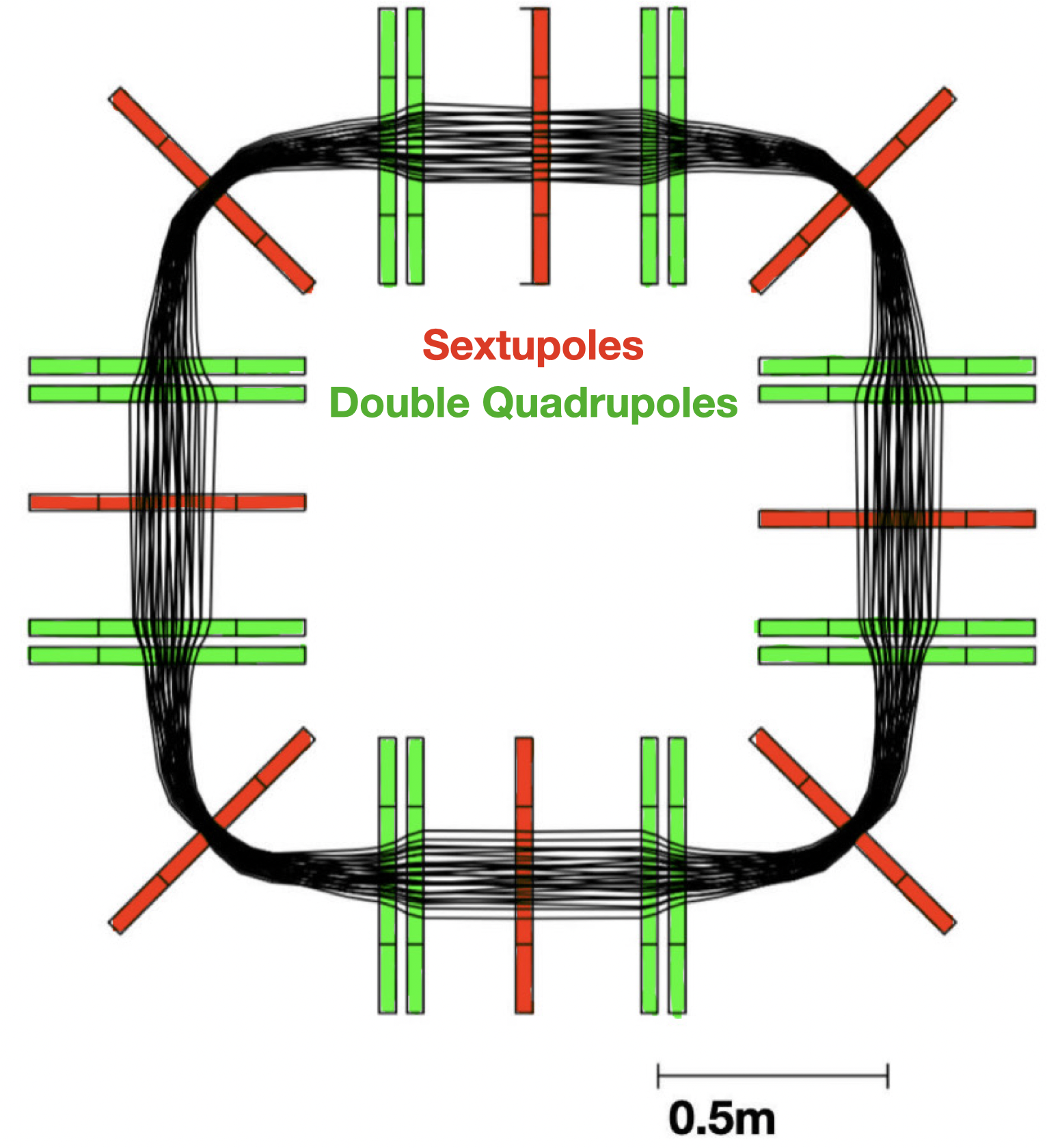}
		\caption{Beam stability simulation with sextupoles and double quadrupoles.}\label{vachousing}
	\end{minipage}
\end{figure}

\noindent The experimental setup comprises an all-electrostatic storage ring for ions and ionic molecules. The demonstrator setup can store up to $10^9$ charged particles in 200 bunches, confined and tuned by electrostatic potentials on periodic trajectories. However, the amount of particles is limited due to space charge effects \cite{spacecharge}. The whole ring is kept at ultra-high vacuum conditions inside a stainless steel housing to enable extremely long trapping times of up to 5~h for the cryogenic setup. Long trapping times for $10^7$ ions were already demonstrated at the CSR in Heidelberg \cite{CSR}, while long spin coherence times have been demonstrated at the COSY ring \cite{COSY}. Access for spectroscopy lasers allows to manipulate and monitor the polarization of the particles during storage, enabling the use of the device as a magnetometer. Through electrostatic optics and radio frequency elements, the particles are bunched and focused such that polarization can be read out either optically or with SQUID magnetometers.

\section{Experimental Setup}
\noindent
 (i) {\em The vacuum system and ion optics}
for the initial implementation will be a demonstrator setup, operated at room temperature, while the final stage will be at cryogenic conditions, similar to the CSR in Heidelberg \cite{CSR}. The chamber comprises deflectors to bend in two steps (6~$^\circ$ and 39~$^\circ$), as well as sextupoles and double quadrupoles for focussing and beam stability reasons as shown in Fig.\ref{sketch} and \ref{vachousing}.
For the demonstrator setup, we first envisage trapping 30~keV barium ions at 36~kHz. The beam stability for these ions was checked with Gicosy \cite{Gicosy}, see Fig. \ref{vachousing}. 
\vspace{0.2cm}
\\
\noindent
\begin{figure}[h]
	\centering
	\includegraphics[scale=0.11]{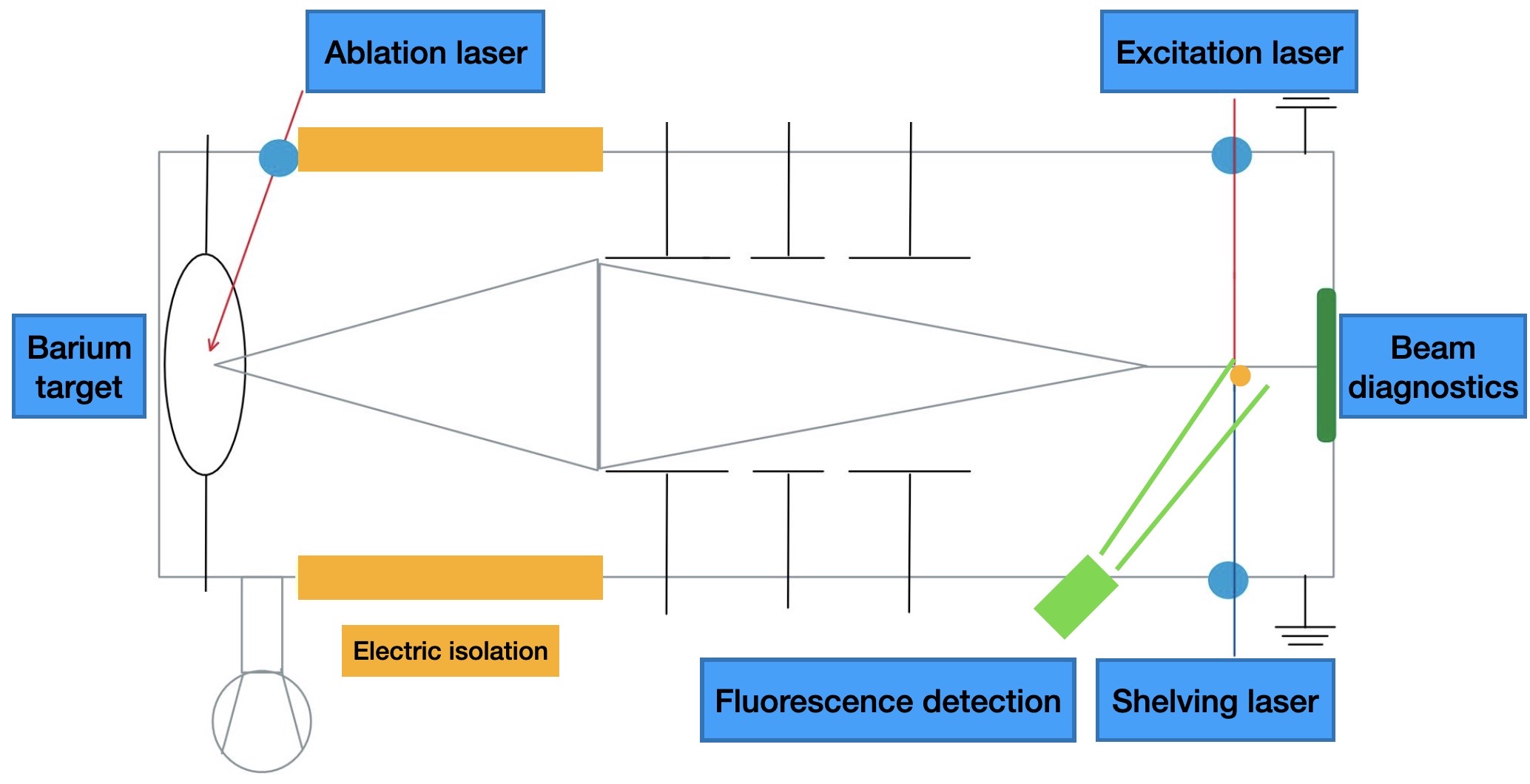}
	\caption{Sketch of the barium ion source vacuum system, consisting of a laser ablation source, an electrostatic isolation, an electrostatic einzellens, 2 lasers for polarization, and a multichannel plate for beam diagnostics.}\label{ionsource}
\end{figure}
(ii) {\em The ion source} is a laser ablation source required to produce $10^6$ ions in 100~ns packets, which are then electrostatically accelerated and injected into the ring. With the laser ablation source, small spot sizes of about $100~\mu m$ or smaller are possible \cite{Murray_2022}. The spot size is monitored with a multi-channel plate. 
To demonstrate the feasibility of such a storage ring, we plan on storing barium ions at first, which have 
no nuclear spin and only one valence electron and can be used to test axion-electron couplings.
 We will be able to set new laboratory limits on axion-electron couplings at the first "switch-on".
 
\vspace{0.2cm}

\noindent
(iii) {\em For polarization} we plan to use a 650~nm laser and a 493~nm laser due to the metastable state \cite{BaSpectroscopy}. The Zeeman splitting of the ground state in earth's magnetic field is 1~MHz, large enough to polarize the ions without further magnetic fields \cite{ZeemanBfield}. We plan on polarizing the ions directly in the ring for the final setup. The feasibility of optical pumping in a storage ring was already observed at the GSI \cite{OpticalPumpingstoragering}, where the observations were explained with rate equations. Adapting these to barium ions, we find that the particles are polarized after about 8s, which fits well within our demonstrator setup, where we can store the ions for up to 1h.
       
\vspace{0.2cm}
\noindent
(iv) {\em Long spin coherence times} for the ions in the storage ring are required to perform spin precession measurements. 
We used a ballistic estimation for the $T_2$ time from polarized ions in a box \cite{SpinRelaxation}.
\begin{equation}
    \frac{1}{T_2}\thicksim \frac{4\gamma^2R^4}{175D}(\nabla B)^2,
\end{equation}
with an estimation for the diffusion constant in the low-pressure limit as
\hbox{$D=v\frac{R^3}{\lambda}=4.17\cdot10^{-3}\frac{m^2}{s}$}. Estimating all contributions to the magnetic field gradients, we arrive at a $T_2$ time of about 21~min. This could be further improved by having a more focused beam and a smaller beam divergence, but it does not limit the first measurements with the demonstrator setup.

\section{Projected sensitivities}
\subsection{EDM measurements with a cryogenic ring}
A naive rescaling from an experiment trying to measure the electron EDM 
using molecular ions confined in a radio frequency trap  \cite{CornellEDM} 
to our setup with $10^9$ particles and 5~h storage time leads to an increased sensitivity of $10^6$ in statistics. This needs to include all possible and expected new systematic issues emerging from transferring a rather sophisticated experimental technique to a novel and quite a different setting and yet unknown difficulties.
\subsection{Axion wind experiment}
\begin{figure}[t]
	\centering
	\includegraphics[width=0.49\linewidth]{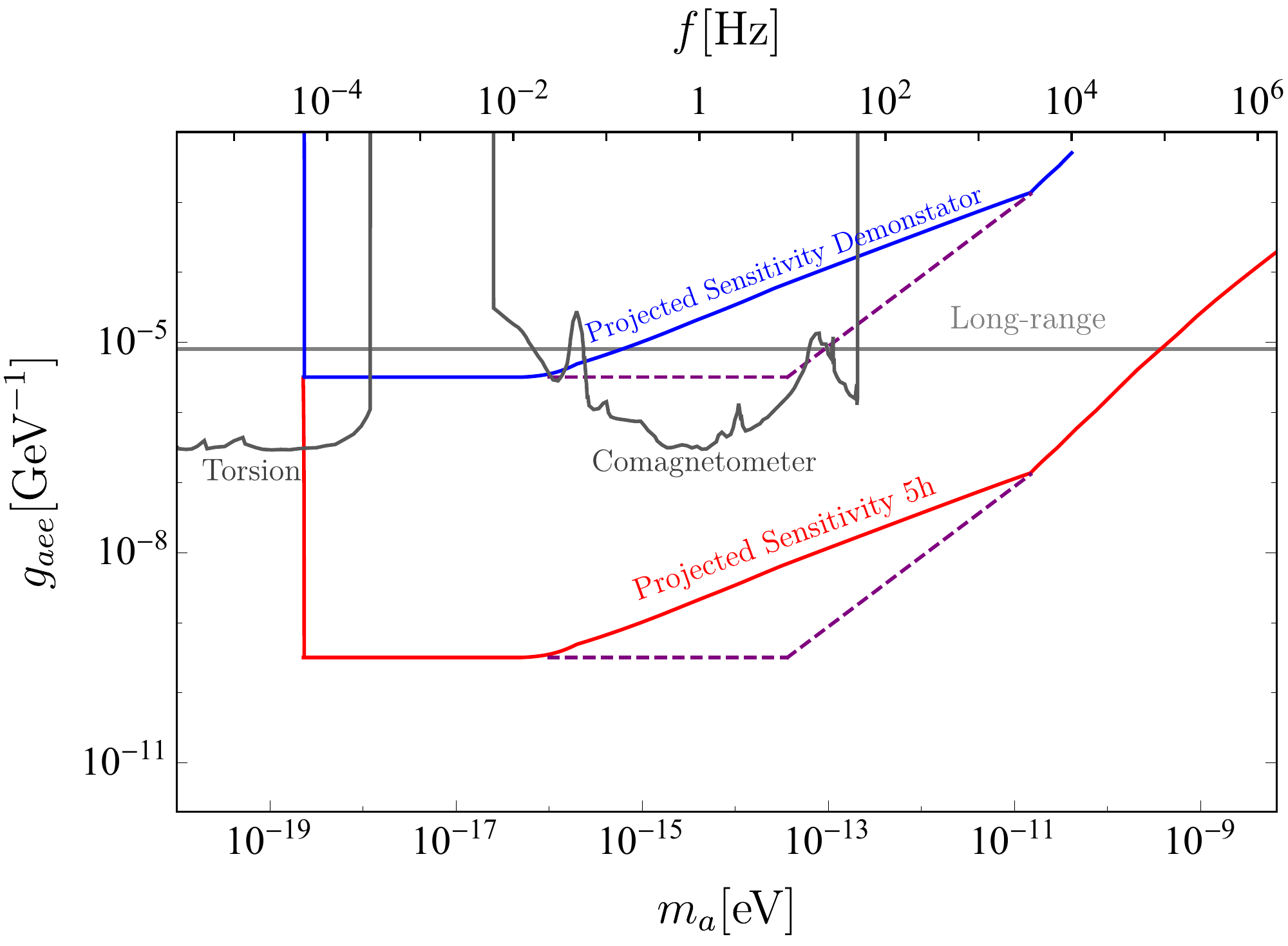}
	\includegraphics[width=0.49\linewidth]{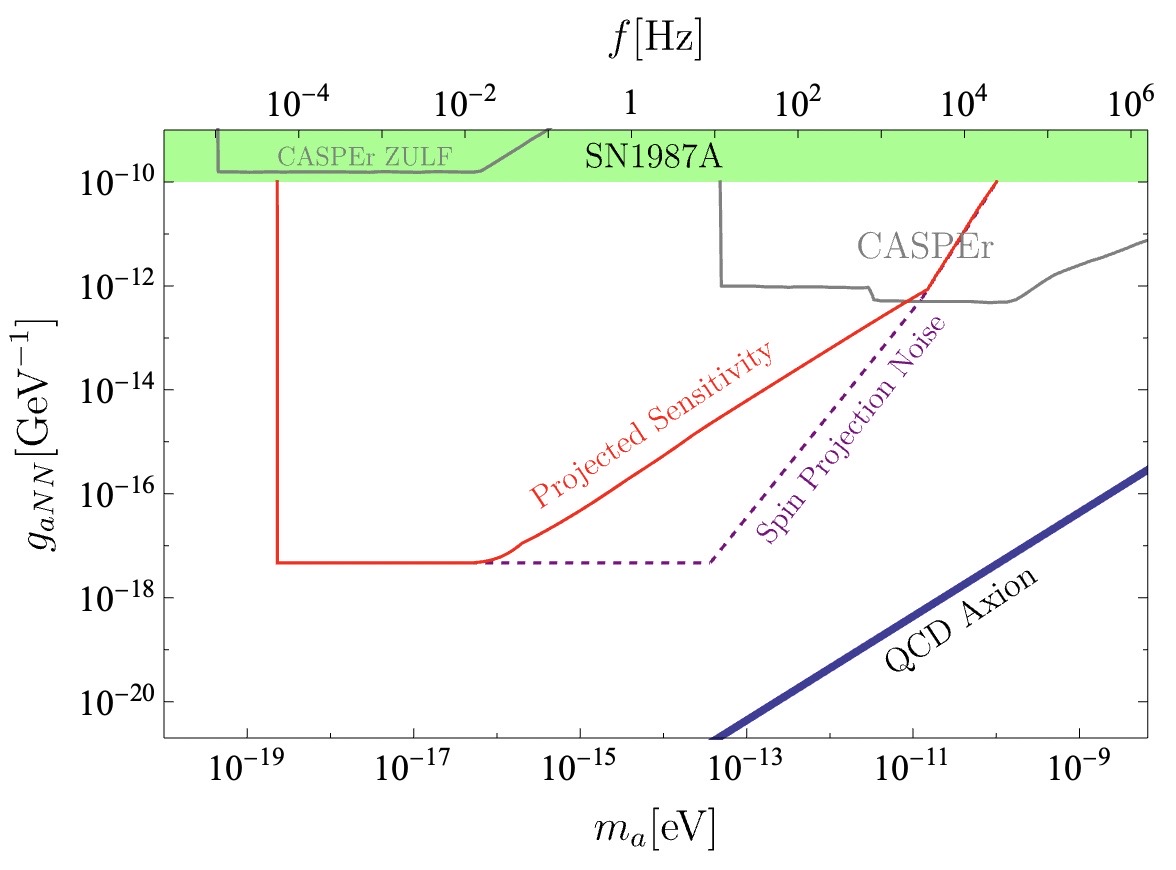}
	\caption{Projected sensitivity of the storage ring. Axion-electron coupling (left), with storage times of 1s (blue) and 5h (red), where we show current bounds from long-range interactions\cite{Terrano:2015sna}, torsion pendulum \cite{Terrano:2019clh} as well as comagnetometer bounds~\cite{axionelectron}. Note that the red line approaches the astrophysical bounds \cite{MillerBertolami:2014rka}. Axion-nucleon couplings (right) for a final stage with 5h storage time and a nuclear spin polarized xenon sample in the center of the ring. The projections of the CASPEr experiment are taken from~\cite{JacksonKimball:2017elr,Garcon:2019inh}.} \label{projectedsensitivity}
\end{figure}

If the axion contributes to the observed dark matter abundance, its couplings to electron and nuclear spins can be interpreted as an oscillating effective background magnetic field \cite{Graham_2021},
\begin{equation}
	\mathcal{L}_{int}\supset g_{a\psi \psi}\partial_\mu a\Bar{\psi}\gamma^5\gamma_\mu \psi \rightarrow H=-g_{a\psi\psi}\nabla a\cdot\sigma.
\end{equation}
This effective magnetic field induces a spin precession in the storage ring, which can be measured either with SQUIDs or with an optical readout. The demonstrator setup can directly probe the axion-electron coupling (Fig. \ref{projectedsensitivity} left panel), while for a final setup, a sample of nuclear spin polarized xenon with density $n\sim 10^{25}\text{m}^{-3}$ and polarization $P\sim0.3$ can be placed in the center of the ring. The projected sensitivity on the axion-nucleon coupling via a double resonant search is shown in Fig. \ref{projectedsensitivity} (right panel). As there is no magnetic shielding of the storage ring, the same search probes kinetically mixed dark photon dark matter \cite{Nelson:2011sf}.

\section*{Conclusion}
A concept for an electrostatic table-top-sized storage ring for polarized particles as a new type of magnetometer is discussed. 
While experimentally challenging, its shot-noise limited sensitivity is $10^{-21}$~T at $<$ mHz bandwidth, making it extraordinarily sensitive for specific applications.

\noindent Among the most exciting use cases is the measurement of effective magnetic fields caused by light-dark matter particles with different types of spin-dependent couplings, which are seen as effective time varying magnetic fields in the experiment.
The magnetometer can also be used as a container for a large number of ionic molecules for electric dipole moment measurements, enabling hours of storage and coherence time with the possibility to control and investigate systematic effects.

\noindent An experimental realization of the first stage of the experiment, including the particle beam and polarization of trapped ions, is currently starting. The barium ion source will be finished beginning next year, and the storage ring vacuum chamber will be set up next year.\\

\noindent This research was supported by the Excellence Cluster ORIGINS, funded by the Deutsche Forschungsgemeinschaft (DFG, German Research Foundation) under Germany’s Excellence Strategy – EXC-2094 – 390783311. The work of SS is additionally supported by the Swiss National Science Foundation under contract 200020-18867.

%
%
%

\end{document}